\documentclass[conference]{IEEEtran}
\IEEEoverridecommandlockouts
\usepackage{cite}
\usepackage{cuted}
\usepackage{amsmath,amssymb,amsfonts}
\usepackage{algorithmic}
\usepackage{graphicx}
\usepackage{textcomp}
\usepackage{xcolor}
\usepackage{verbatim}
\def\BibTeX{{\rm B\kern-.05em{\sc i\kern-.025em b}\kern-.08em
    T\kern-.1667em\lower.7ex\hbox{E}\kern-.125emX}}
\begin{document}

\title{Graphical State Space Model}% and Its Application in Integrated Navigation\\

\author{\IEEEauthorblockN{ShaoLin L{\"u}}%\textsuperscript{*}, YiHong Ge}
\IEEEauthorblockA{
\textit{GraphOptimization Inc.}\\
Beijing, China \\
4050627@qq.com}
}

\maketitle

\begin{abstract}
In this paper, a new framework, named as graphical state space model,  is proposed
for the real time optimal estimation of one kind of nonlinear state space model.
By discretizing this kind of system model as an
equation which can not be solved by Extended Kalman filter, factor graph optimization can outperform Extended
Kalman filter in some cases.
A simple nonlinear example is given to demonstrate the efficiency of this framework.
\end{abstract}


\section{Introduction}
Optimal estimation\cite{b1}-\cite{b4} is a hit subject for engineers, especially in the science community of GPS, inertial navigation or integrated navigation.
In many fields, Kalman filter\cite{b5} is the chosen one in the eyes of countless engineers.
Recently, factor graph\cite{b6}-\cite{b7} was frequently used to solve many optimal estimation problems in robotics community. It has been widely admitted that  on
SLAM, SFM and many other subjects better results can be achieved by factor graph optimization than by Kalman filter.
However, these good results  usually emerged in the cases of post processing. For real time processing, Kalman filter still
occupies the hearts of numerous engineers.
\par In the last forty years, Probabilistic Graphical Model\cite{b8} became popular in both  academic world and industry world. Pearl\cite{b9} introduced the belief propogation to Bayesian network, which can solve the propability inference problem
 in complex graphical models. Loeiger\cite{b7} discovered sum-product algorithm for factor graph, which motivated great improvements in communication community.
 Dellaert and Kaess\cite{b6} developed gorgeous techniques for
 factor graph, while numerous applications  in the robotics community were stimulated by their research.
 \par
 Generally, there are two kinds of graphcial models\cite{b8}: Bayesian network\cite{b9} (Directional Graphical Model) and Markov Field (Undirectional Graphical Model).
 Both of them can be transformed to factor graph. Then, powerful tools, such like junction tree or other appoximate inference algorithms can be used to
 perform  probabilistic inference on factor graph.
 According to the pioneer work mentioned above,
 we have enough tools to  solve much more complex graphical models than those solved by the pioneers in 1960s.
 Kalman filter is regarded as a special simple case of dynamic Bayesian network\cite{b10} from the viewpoint of graphical models.
\par In this paper, a new framework, named as Graphical State Space Model(GSSM), is proposed for the real time optimal estimation of one kind of nonlinear state space model.
In the framework of Kalman filter, the continous-time state space model is discretized as a time-series model in the chronological order. Here
 it is demonstrated that some kind of continous-time model can be discretized as a dynamic Bayesian network.
  Then, factor graph can be used to represent the density functions. This provides the possibility in
 the increase of the estimation accuracy.
Another merit of this dicretization method is that it can reduct the dimension of the equation needed to be solved.  A simple example is used to demonstrate the availibility of the new framework. The clou of this paper is how to organize an equation that Kalman filter can not solve while factor graph optimization can solve.
\par The paper is orginized as follows. Section \uppercase\expandafter{\romannumeral1} is a brief introduction.
In section \uppercase\expandafter{\romannumeral2}, Kalman filter is analyzed from the viewpoint of graphical model.
In section \uppercase\expandafter{\romannumeral3}, it is demonstrated  what factor graph optimization can do for standard discrete state space model.
Grapical state space model is proposed in section \uppercase\expandafter{\romannumeral4}. A simple example is given to demonstrate the availibility of this powerful model in section \uppercase\expandafter{\romannumeral5}.
This paper is concluded in section \uppercase\expandafter{\romannumeral6}.
\section{Graphical model viewpoint for Kalman Filter}

%\subsection{Maintaining the Integrity of the Specifications}
There are many viewpoints for Kalman filter, such as Bayesian filter, information filter or recursive weighted least squared methods.
 Before we investigate Kalman filter at
the perspective of graphical model, let us review this classical algorithm in the traditional manner.
\par
Besides the famous equations of Kalman filter, the prequisite part is continous-time system theory and its discretization.
 Many problems in the real world can be formulated as a nonlinear system.
\begin{gather}
\dot{x}=f(x)+B(u)+q\\
y=C(x)+r
\end{gather}
The linearization of the above system  can be discribed as follows
\begin{gather}
\dot{x}=Ax+Bu+q\\
y=Cx+r
\end{gather}
where $x\in \bold{R}^n $ is the system state vector, $y\in \bold{R}^m $ is the measurement vector,   $u\in \bold{R}^l $ is the control vector.
$A$ is an $n \times n$ matrix, which is called  the system matrix.   $B$ is an $n \times l$ matrix, which is called  the input matrix.
$C$ is an $m \times n$ matrix, which is called  the measurement matrix. $q \sim N(0,Q)$ is the process noise and $r\sim N(0,R)$ is the measurement noise.
$Q$ and $R$ are the known covariance matrices.

\par
The above system can be discretized as follows
\begin{gather}
x_{k+1}=F_kx_k+B_ku_k+q_k, F_k\approx I_{n \times n}+AT\\
y_{k+1}=C_{k+1}x_{k+1}+r_k
\end{gather}
where $x_k\in \bold{R}^n $ is the discrete system state vector, $y_{k+1}\in \bold{R}^m $ is  the discrete measurement vector,   $u_k\in \bold{R}^l $ is the discrete control vector.
$F_k$ is an $n \times n$ matrix, which is called  the  system matrix or the predition matrix. $T$ is the sample time interval.
$B_k$ is an $n \times l$ matrix, which is called  the  input matrix.
$C_{k+1}$ is an $m \times n$ matrix, which is called  the  measurement  matrix. $q_{k}\sim N(0,Q_k)$ is the process noise and $r_k\sim N(0,R_k)$ is the measurement noise. $Q_k$ and $R_k$ are the known covariance matrices.
\par
The classical Kalman filter can be used to estimate the system  described by Equations $(5) \sim (6)$.\\
Initialize the  estimation,
\begin{gather}
\hat{x}_{0}=E(x_0)=x_0^+ \\
P_0=E[(x_0-\hat{x}_{0})(x_0-\hat{x}_{0})^T]
\end{gather}
Update the prior mean and covariance
\begin{gather}
x_{k+1|k}=F_kx_k+B_ku_k\\
P_{k+1|k}=F_kP_kF_k^T+Q_k
\end{gather}
Calculate Kalman gain
\begin{gather}
K_{k+1}=P_{k+1|k}C_{k+1}^T(C_{k+1}P_{k+1|k}C_{k+1}^T+R_k)^{-1}
\end{gather}
Update the posteriori mean and covariance
\begin{gather}
x_{k+1}=x_{k+1|k}+K_{k+1}(y_{k+1}-C_{k+1}x_{k+1|k})\\
P_{k+1}=(I-K_{k+1}C_{k+1})P_{k+1|k}
\end{gather}

To summarize, Kalman filter consists of three steps:\\
\uppercase\expandafter{1}. Utilize the physical laws to model the system correctly, which goes as Equations  $(1)\sim (4)$.\\
\uppercase\expandafter{2}. Disretize the system, which goes as Equations  $(5)\sim (6)$. \\
\uppercase\expandafter{3}. Using the famous Equations $(7)\sim (13)$ to update the mean and covariance of  $ x_k $ in the chronological order. From the viewpoint of graphical models, Kalman filter can be ragarded as  a specical case of dynamic Bayesian network.
\begin{figure}[htbp]
\centerline{\includegraphics[scale=0.32]{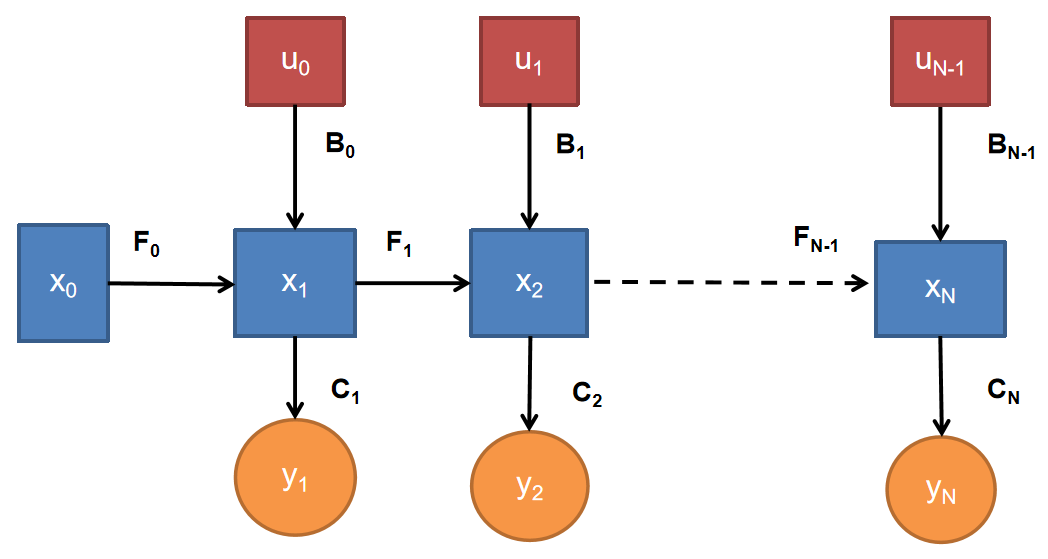}}
\caption{State space model representation of Kalman filter.}
\label{fig}
\end{figure}
\begin{figure}[htbp]
\centerline{\includegraphics[scale=0.28]{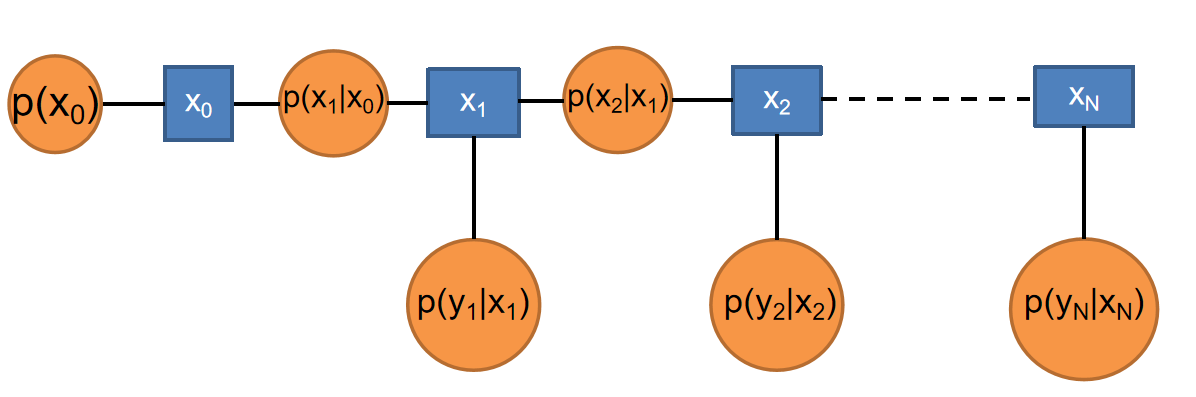}}
\caption{Factor Graph Representation of Kalman filter.}
\label{fig}
\end{figure}

\par Probabilistic graphicl model theory consists of three parts\cite{b8}: representation, inference and learning. In Fig. 2, the optimization problem is represented as a factor graph.
Factor graph \cite{b6} is a kind of bipartie graph, $ F=( \bold{U}, \bold{V}, \bold{E}) $ , which consists of Factor nodes, Variable nodes and Edges.
 When inference is done via  probabilistic graphicl model theory, Hidden Markov Model assumption \cite{b10} for  density function is accepted as follows
\begin{gather}
p(x_{0:N},y_{1:N})=p(x_0)\prod\limits_{k=1}^{N}p(x_k|x_k-1)p(y_k|x_k)
\end{gather}
\par In fact, what Kalman filter do in the process of solving Equation (14) is recursively solving a large block tridiagonal equation as follows
\begin{gather}
    P_w=diag
\begin{bmatrix}
P_{0}\\
Q_{0}\\
R_{1}\\
Q_{1}\\
R_{2}\\
\vdots\\
Q_{N-1}\\
R_{N}\\
\end{bmatrix},\notag
\\
\begin{bmatrix}
I_{n\times n} & 0_{n\times n} & 0_{n\times n}&\cdots & 0_{n\times n} & 0_{m\times n}\\
-F_{0} & I_{n\times n}& 0_{n\times n}&\cdots & 0_{n\times n} & 0_{n\times n}\\
0_{m\times n} & C_{1}& 0_{m\times n}&\cdots & 0_{m\times n} & 0_{m\times n}\\
0_{n\times n}&-F_{1} & I_{n\times n}& \cdots & 0_{n\times n} & 0_{n\times n}\\
0_{m\times n} & 0_{m\times n}& C_{2}&\cdots & 0_{m\times n} & 0_{m\times n}\\
\vdots &\vdots &\vdots &\ddots & \vdots & \vdots \\
0_{n\times n} & 0_{n\times n} & 0_{n\times n}& \cdots &-F_{N-1} & I_{n\times n}\\
0_{m\times n} & 0_{m\times n}& 0_{m\times n}&\cdots & 0_{m\times n} & C_{N}
\end{bmatrix}\notag
\\
\begin{bmatrix}
x_0  \\
x_{1}\\
x_{2}\\
\vdots \\
\vdots \\
x_{N-1}\\
x_{N}
\end{bmatrix}
=
\begin{bmatrix}
\hat{x}_0^+  \\
B_0u_0\\
y_{1}\\
B_1u_1\\
y_{2}\\
\vdots \\
B_{N-1}u_{N-1}\\
y_{N}\\
\end{bmatrix}
\end{gather}
\par At time $ k+1 $, probabilistic inference is done on a small factor graph represented by Fig. 4.
\begin{figure}[htbp]
\centerline{\includegraphics[scale=0.34]{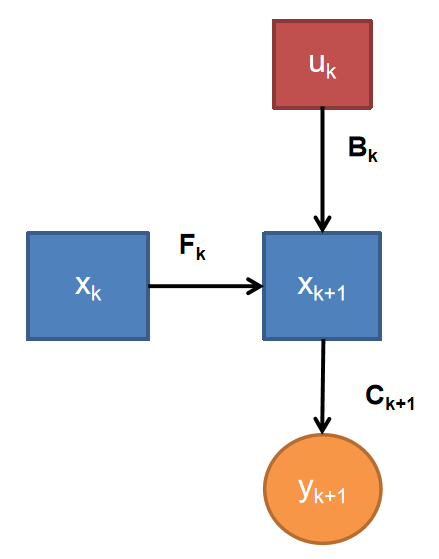}}
\caption{ State space model representation  of Kalman filter at time $ k+1 $.}
\label{fig}
\end{figure}
\begin{figure}[htbp]
\centerline{\includegraphics[scale=0.28]{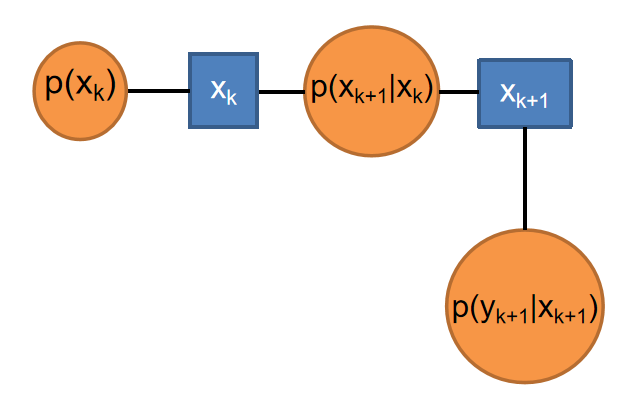}}
\caption{Factor Graph representation of Kalman filter at time $ k+1 $.}
\label{fig}
\end{figure}

\par
According to message-passing algorithm theory\cite{b10}, Kalman filter can only perform forward directional message passing .
The reason is that marginalization, which is discribed by Equations $(12)\sim (13)$,  is done  every each time.
This means that Kalman filter can not pass the later messages backward to the previous states.
\par The inference problem represented by Fig. 2 can be perfermed by the solution of the equation as follows
\begin{small}
\begin{gather}
      P_w=diag
\begin{bmatrix}
P_{k}\\
Q_{k}\\
R_{k+1}
\end{bmatrix},
\begin{bmatrix}
I_{n\times n} & 0_{n\times n}\\
-F_{k} & I_{n\times n}\\
0_{m\times n} & C_{k+1}
\end{bmatrix}
\begin{bmatrix}
x_k  \\
x_{k+1}
\end{bmatrix}
=
\begin{bmatrix}
\hat{x}_k  \\
B_ku_k\\
y_{k+1}
\end{bmatrix}
\end{gather}
\end{small}
\par Kalman filter utilizes Equations $(9)\sim (13)$ to solve Equation (16). Different from factor graph optimization, Kalman filter only give the
estimation of $x_{k+1}$, while factor graph optimization will calculate the estimation of  $x_{k+1}$ and  $x_{k}$ at the same time.
\par For factor graph optimization, the above equation can be arranged as three factors \cite{b6}:\\
\uppercase\expandafter{1}. Priori factor, which represents the priori $p(x_k)$.\\
\uppercase\expandafter{2}. Between factor, which represents the transition probability $p(x_{k+1}|x_k)$.\\
\uppercase\expandafter{3}. Measurement factor, which represents the conditional probability $p(y_{k+1}|x_{k+1})$.
\begin{gather}
p(x_{k},x_{k+1}|y_{k+1})\propto p(x_k)p(x_{k+1}|x_k)p(y_{k+1}|x_{k+1})
\end{gather}
\par The junction tree can be generated for these factors.
 The clique is simplely comprised of three factors above. Via this tree, probabilistic inference is performed.
 From the viewpoint of factor graph optimization, the distributions of $x_k $ and $x_{k+1} $ are updated while at
 the perspective of Kalman filter, only the distribution of $x_{k+1}$ is updated, which is called marginalization.
\par Obviously, Kalman filter is a specical case of factor graph optimization. First, Kalman filter have no iteration steps.
Secondly, Kalman filter does not update the distributions of previous states, which is called smoothing at the perspective of traditional optimal estimation.

\section{Factor Graph Optimization for State Space model}
\par It should be noted that Equation (15) can be solved easily be factor graph optimization. Assuming that
\begin{gather}
p(x_0)\propto exp\{-\frac{1}{2}\|x_0-x_0^+\| _{P_0} \} \\
p(x_k+1|x_k)\propto exp\{-\frac{1}{2}\|x_{k+1}-F_{k}x_{k}-B_ku_k\| _{Q_k} \}\\
p(y_{k+1}|x_{k+1}\propto exp\{-\frac{1}{2}\|y_{k+1}-C_{k}x_{k+1}\| _{R_k+1} \}
\end{gather}
 \par Then, Equation (15) can be  formulated as a weighted least squared problem
 \begin{small}
 \begin{gather}
\bold{X}^{MAP}={\mathop{argmin} \limits_{\bold{X}}
{( \|x_0-x_0^+\| _{P_0} }}
+{\mathop{\sum} \limits_{k}( \|x_{k+1}-F_{k}x_{k}-B_ku_k\| _{Q_k}}
\notag
\\
+ \|y_{k+1}-C_{k}x_{k+1}\| _{R_{k+1}}))
\end{gather}
\end{small}
 where $\bold{X}=\{ x_k \},k=0,...,N$.
 \par
 Open source library,
such as Ceres\cite{b11}, G2O\cite{b12}, GTSAM\cite{b6}, miniSAM\cite{b13} and minisam\cite{b14}, can be easily gotten via internet to solve the above problem. In fact, factor graph optimization can solve both single connected graphical model, such as
state space model, and multi connected graphical model. It is a powerful bottle-opener for complex graphical models.

There are two kinds of methods for factor graph optimization: global optimization and sliding window optimization. The difference lies in whether do marginalization or not.
Since the dimension of the whole graph increases as time goes on, the amounts of caculation will be out of control if the dimension of the graph is not limited.
For this reason, when factor graph optimization is used in real-time processing case, sliding window optimization will be adopted, which  can be represented as Fig. 5 and Fig. 6.
\begin{figure}[htbp]
\centerline{\includegraphics[scale=0.32]{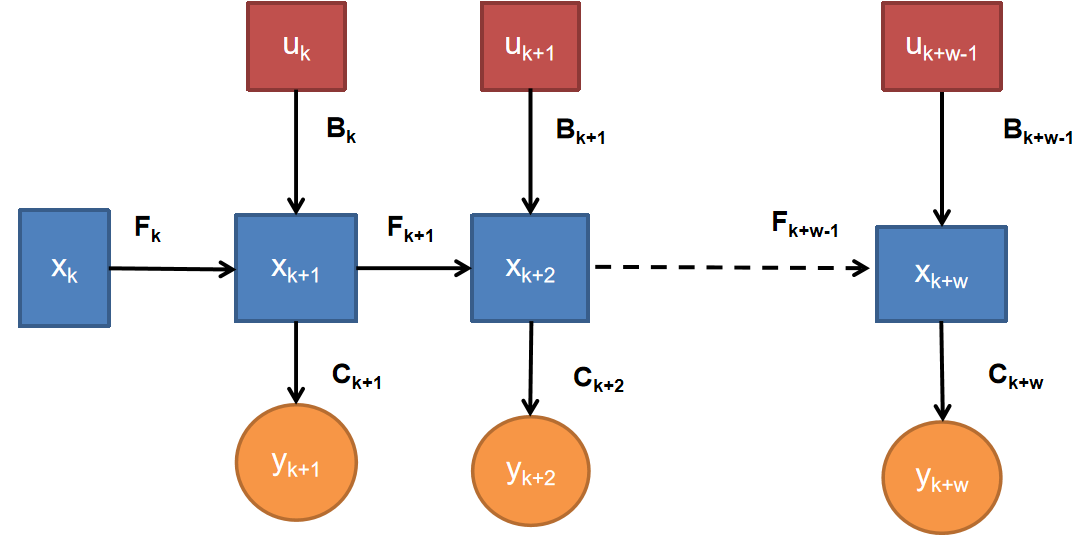}}
\caption{State Space model in Sliding window case.}
\label{fig}
\end{figure}

\begin{figure}[htbp]
\centerline{\includegraphics[scale=0.28]{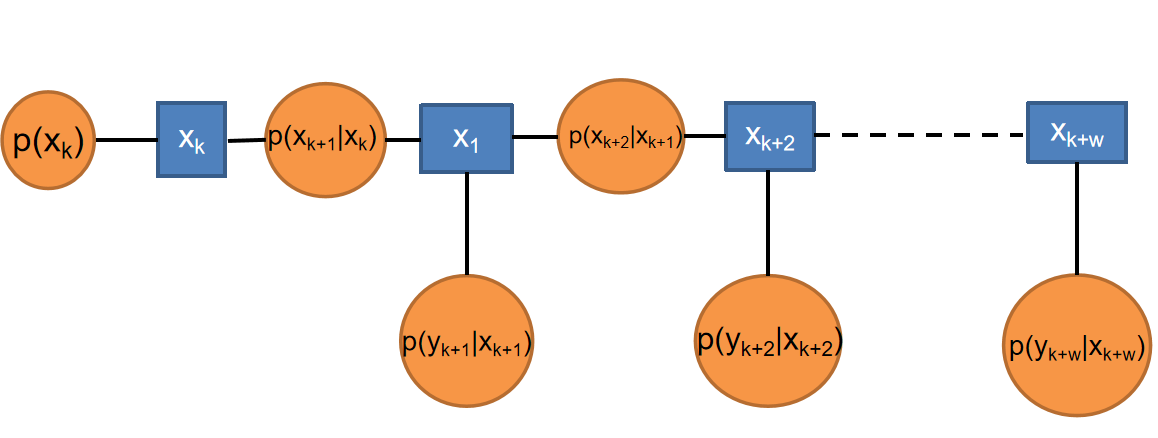}}
\caption{Factor Graph Representation in Sliding window case.}
\label{fig}
\end{figure}
\par The block tridiagonal equation goes as follows
\begin{gather}
  P_w=diag
\begin{bmatrix}
P_{k}\\
Q_{k}\\
R_{k+1}\\
Q_{k+1}\\
R_{k+2}\\
\vdots\\
Q_{k+w-1}\\
R_{k+w}\\
\end{bmatrix},
X_w=
\begin{bmatrix}
x_k  \\
x_{k+1}\\
x_{k+2}\\
\vdots \\
\vdots \\
x_{k+w-1}\\
x_{k+w}
\end{bmatrix},\notag
\\
b_w=
\begin{bmatrix}
\hat{x}_k^+  \\
B_ku_k\\
y_{k+1}\\
B_{k+1}u_{k+1}\\
y_{k+2}\\
\vdots \\
B_{k+w-1}u_{k+w-1}\\
y_{k+w}\\
\end{bmatrix}, \notag
\\
A_w=
{\small
\begin{bmatrix}
I_{n\times n} & 0_{n\times n} & 0_{n\times n}&\cdots & 0_{n\times n} & 0_{n\times n}\\
-F_{k} & I_{n\times n}& 0_{n\times n}&\cdots & 0_{n\times n} & 0_{n\times n}\\
0_{m\times n} & C_{k+1}& 0_{m\times n}&\cdots & 0_{m\times n} & 0_{m\times n}\\
0_{n\times n}&-F_{k+1} & I_{n\times n}& \cdots & 0_{n\times n} & 0_{n\times n}\\
0_{m\times n} & 0_{m\times n}& C_{k+2}&\cdots & 0_{m\times n} & 0_{m\times n}\\
\vdots &\vdots &\vdots &\ddots & \vdots & \vdots \\
0_{n\times n} & 0_{n\times n} & 0_{n\times n}& \cdots &-F_{k+w-1} & I_{n\times n}\\
0_{m\times n} & 0_{m\times n}& 0_{m\times n}&\cdots & 0_{m\times n} & C_{k+w}
\end{bmatrix}}
\notag
\\
A_wX_w=b_w
\end{gather}
\begin{gather}
\hat{X}_w=(A_w^TP_w^{-1}A_w)^{-1}A^T_wP_w^{-1}b_w
\end{gather}
\par After the above equation is solved, $x_{k+1}$ should be marginalized out for next measurement update.
\par
This sliding window model just has a bigger dimension than the model discribed in Fig. 4. When factor graph optimization is used to solve Equation (22),
its solution accuracy is just like the sliding window filter in theory.
It does not use another important merit of factor graph, which is that factor graph can model the system into a multi connected graph.
\par
It should not be igored, what factor graph can do while the traditional  space model can not do:
\\
\uppercase\expandafter{1}. Model the system as a multi connected graph.\\
\uppercase\expandafter{2}. Model the system  distributedly not unifiedly.\\
\uppercase\expandafter{3}. Model one part of the system  as constants rather than time-series variables.
\par These merits are not utilized in Equation (22).
If a new framework can be designed to adopt them, maybe factor graph optimization can outperform Kalman filter in real-time processing case.

\section{Graphical state  space model}
It is strongly believed by many engineers that Extended Kalman filter is the best solution for sequential data optimal estimation described by Equations $(1)\sim (6)$.
We totally agree with this opinion. Although graphical models solved  by factor graph optimization can be  much more complex  than  space model, factor graph optimization can not outperform
 Kalman filter when they use the same sequential space model.
 Another fact which can not be ignored is that factor graph  can model the system  distributedly. When Kalman filter is used to model the system,
 there is only one universial state which evolves in the chronological order.
 If you want better results via factor graph optimization, perhaps you should try a different state space model.
\par For some continous-time system decribed by Equations $(3)\sim (4)$, there is a different discretization method.
For this kind of system, it can be discretized as another kind of graphical model. Consider the system as follows
\begin{gather}
  \begin{bmatrix}
  \dot{x_c}  \\
  \dot{x_b}
  \end{bmatrix}
=
\begin{bmatrix}
A_c & A_{b} \\
0 & 0
\end{bmatrix}
\begin{bmatrix}
x_c \\
x_b
\end{bmatrix}
+
\begin{bmatrix}
B \\
0
\end{bmatrix}
u+
\begin{bmatrix}
q \\
0
\end{bmatrix} \\
y=
\begin{bmatrix}
 C_c & C_b
 \end{bmatrix}
 \begin{bmatrix}
  x_c  \\
  x_b
  \end{bmatrix}+r
\end{gather}
where   $ [ x_c, x_b  ] ^T $ is the system state vector, $x_c\in \bold{R}^{n_c}, x_b\in \bold{R}^{n_b}, n_c+n_b=n$.  $y\in \bold{R}^m $ is the measurement vector,   $u\in \bold{R}^l $
is the control vector.
$A_c$ is an $n_c \times n_c$ matrix and $A_b$ is an $n_c \times n_b$ matrix.   $B$ is an $n_c \times l$ matrix, which is called  the input matrix.
$C_c$ is an $m \times n_c$ matrix and $C_b$ is an $m \times n_b$ matrix. $q \sim N(0,Q_c)$ is the process noise and $r\sim N(0,R)$ is the measurement noise.
$Q$ and $R$ are the known covariance matrices.

\par When Kalman filter is used to estimate the state, $ [ x_c, x_b  ] ^T  $, in the above equations, Equations $(5)\sim (6)$
will be used to discretize Equations  $(24)\sim (25)$. The prediction matrix will be
\begin{gather}
F_k\approx I_{n \times n}+AT\notag
\\
A=\begin{bmatrix}
A_c & A_{b} \\
0 & 0
\end{bmatrix}
\end{gather}
\par Next, Kalman filter will recursively solve Equation (15). Obviously, Kalman filter can not make use of the characteristics of this kind of this system.
\par Graphical state space model is different with the  system model used by Kalman filter. No matter in global or sliding window case, it discretizes the system as a graphical model which has a different topological structure.
\par
The  system described by Equations $(24)\sim (25) $ can be discretized as follows
\begin{gather}
x_c(k+1)=F_c(k)x_c(k)+F_b(k)x_b+B_ku_k+q_c(k)\notag
\\
F_c(k)\approx I_{n \times n}+A_cT,
F_b=A_bT\\
y_{k+1}=C_c(k+1)x_c(k+1)+C_b(k+1)x_b+r_k\\
E(x_b)=x_b^+
\end{gather}
where $x_c(k)\in \bold{R}^{n_c}, x_b\in \bold{R}^{n_b}, n_c+n_b=n$.  $y_{k+1}\in \bold{R}^m $ is  the  measurement vector,   $u_k\in \bold{R}^l $ is the  control vector.
$F_c(k)$ is an $n_c \times n_c$ matrix and $F_b(k)$ is an $n_c \times n_b$ matrix. $T$ is the sample time interval.
$B_k$ is an $n_c \times l$ matrix, which is called  the  input matrix.
$[C_c(k+1) \ C_b(k+1)]$ is an $m \times n$ matrix, which is called  the  measurement  matrix. $q_c(k)\sim N(0,Q_k)$ is the process noise and $r_k\sim N(0,R_k)$ is the measurement noise. $Q_k$ and $R_k$ are the known covariance matrices.
\par Equation (29) is the dicretization of the state $x_b$. This means that constant states can be modeled as a different form rather than a time-series form. This is the distributed discretization not the unified discretization.
\par Definitely, Kalman filter can not solve Equations $(27)\sim (29) $ if unknown $x_b $ is not discretized as a time-series form.
This is the limitation of Kalman filter that it  must use time-series forms to represent constant variables.
Via the kind of discretization described in Equations  $(27)\sim (29) $ , the equation  solved by factor graph optimization will be different with Equation (22).
\begin{figure}[htbp]
\centerline{\includegraphics[scale=0.28]{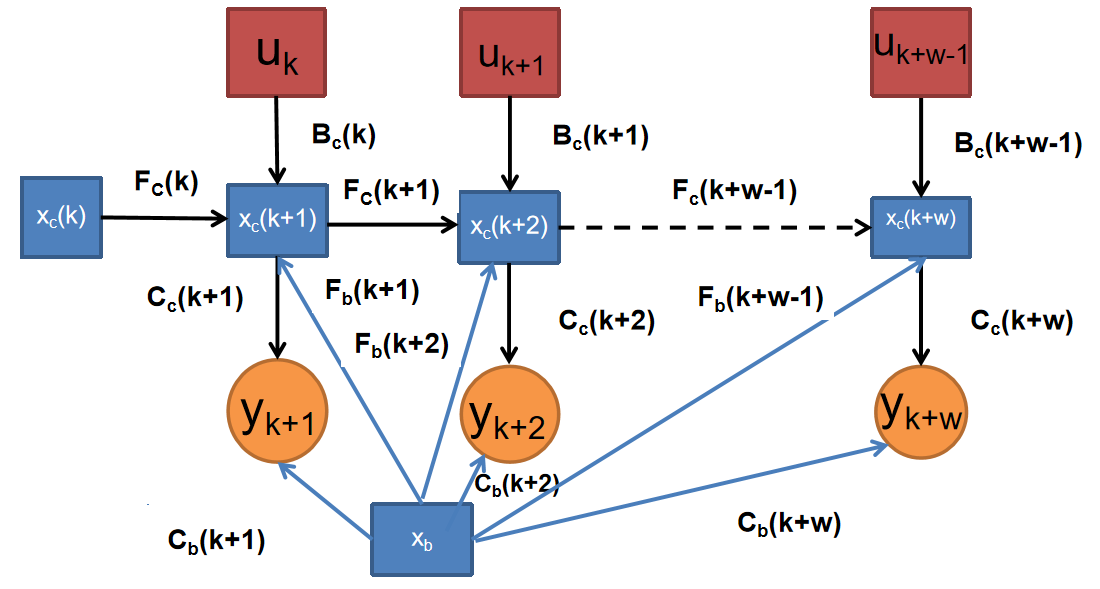}}
\caption{Graphical state space model represented in state space form for sliding window optimization case.}
\label{fig}
\end{figure}
\begin{figure}[htbp]
\centerline{\includegraphics[scale=0.28]{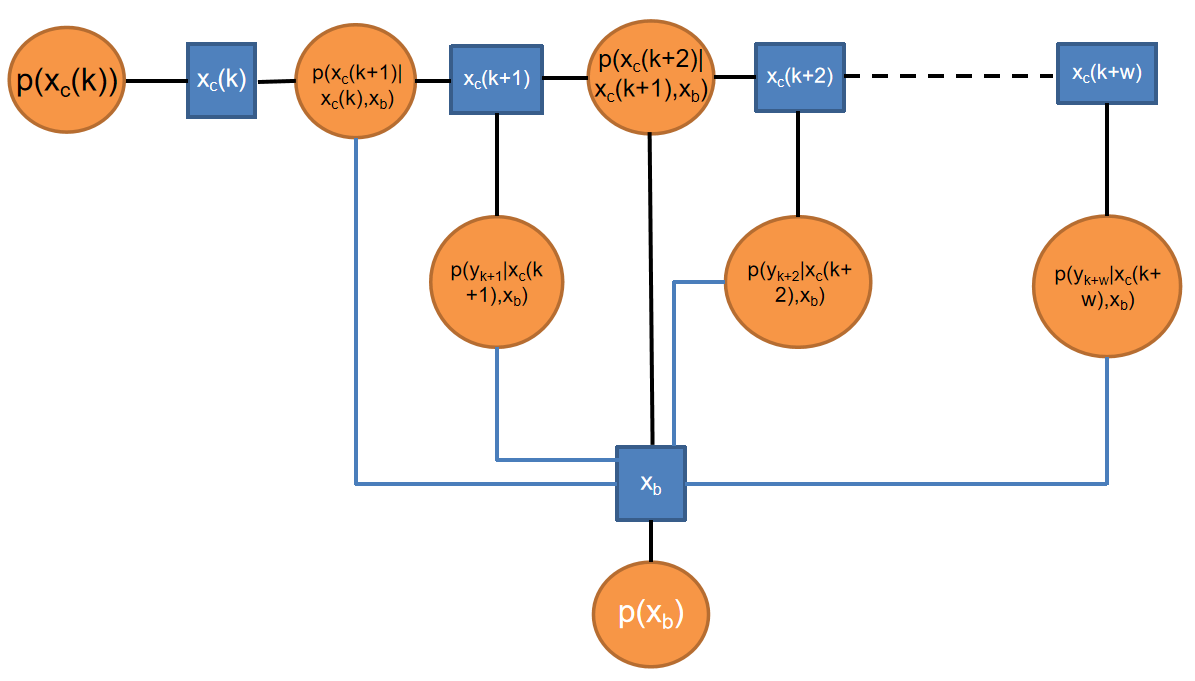}}
\caption{Graphical state space model represented in factor graph form for sliding window optimization case.}
\label{fig}
\end{figure}
\par
Furthermore, since
$x_b$ is not discretized as a time-series form, the dimension of the equation is smaller than that of Equation (22).
The variables are ${x_b,{x_c(i)},i=k,...,k+w}$, while if the discretization method of Kalman filter is used, the variables of Equation (22) will be ${{x_b(i),x_c(i)},i=k,...,k+w}$.
Equation (30) can not be solved by Kalman filter, while factor graph optimization can solve it efficiently.
Since the structure of $A_w$ is not block tridiagonal any more, Kalman filter can not be used recursively to  solve the equation.
By the way, the norm function are smaller than that discribed by Equation (21).
This makes it possible that factor graph optimization can achieve different results with Kalman filter in real-time cases.
\begin{figure*}[ht]
\begin{align}
A_w=
\begin{bmatrix}
I_{n_b\times n_b} & 0_{n_b\times n_c} & 0_{n_b\times n_c} & 0_{n_b\times n_c}&\cdots & 0_{n_b\times n_c} & 0_{n_b\times n_c}\\
0_{n_c\times n_b} & I_{n_c\times n_c} & 0_{n_c\times n_c} & 0_{n_c\times n_c}&\cdots & 0_{n_c\times n_c} & 0_{n_c\times n_c}\\
-F_b(k) & -F_c(k)& I_{n_c\times n_c} & 0_{n_c\times n_c}&\cdots &0_{n_c\times n_c} & 0_{n_c\times n_c}\\
C_b(k+1) & 0_{m\times n_c}& C_c(k+1)& 0_{m\times n_c}&\cdots & 0_{m\times n_c} & 0_{m\times n_c}\\
-F_b(k+1)&0_{n_c\times n_c}& -F_c(k+1)& I_{n_c\times n_c} &\cdots & 0_{n_c\times n_c} & 0_{n_c\times n_c}\\
C_b(k+2) & 0_{m\times n_c} &  0_{m\times n_c} & C_c(k+2) &\cdots &  0_{m\times n_c} & 0_{m\times n_c}\\
\vdots &\vdots &\vdots  &\vdots  &\ddots & \vdots & \vdots \\
-F_b(k+w-1) & 0_{n_c\times n_c}& 0_{n_c\times n_c} & 0_{n_c\times n_c} & \cdots &-F_c(k+w-1) & I_{n_c\times n_c}\\
C_b(k+w) & 0_{m\times n_c} &0_{m\times n_c} &0_{m\times n_c}&\cdots &0_{m\times n_c} & C_c(k+w)
\end{bmatrix} \notag
\end{align}
\end{figure*}

\begin{gather}
  P_w=diag
\begin{bmatrix}
P_{b}\\
P_{c}(k)\\
Q_c(k)\\
R_{k+1}\\
Q_c(k+1)\\
R_{k+2}\\
\vdots\\
Q_c(k+w-1)\\
R_{k+w}\\
\end{bmatrix},
X_w=
\begin{bmatrix}
x_b  \\
x_c(k)\\
x_c(k+1)\\
x_c(k+2)\\
\vdots \\
\vdots \\
x_c(k+w-1)\\
x_c(k+w)
\end{bmatrix},
\notag
\end{gather}
\begin{gather}
b_w=
\begin{bmatrix}
\hat{x}_b^+  \\
\hat{x}_c^+(k)  \\
B_ku_k\\
y_{k+1}\\
B_{k+1}u_{k+1}\\
y_{k+2}\\
\vdots \\
B_{k+w-1}u_{k+w-1}\\
y_{k+w}\\
\end{bmatrix},\notag
\\
A_wX_w=b_w
\end{gather}
\par Via factor graph optimization,  the solution of Equation (30) can be gotten
\begin{gather}
\hat{X}_w=(A_w^TP_w^{-1}A_w)^{-1}A^T_wP_w^{-1}b_w
\end{gather}
\par It is easy to be caculated that the  dimension of the equation is smaller than that dimension if
sliding window Kalman filter was used.

\begin{table}[htbp]
\caption{Dimension Comparision}
\begin{center}
\begin{tabular}{|c|c|c|c|}
\hline
\textbf{Viariable} & \textbf{Dim of $A_w$}& \textbf{Dim of $b_w$}& \textbf{Dim of $X_w$} \\
\hline
\textbf{Sliding window } & $(n_b+nc+m)w\times$  & $(n_b+$ & $(n_b+n_c)$ \\
\textbf{Kalman filter} & $(n_b+n_c)(w+1)$  & $n_c+m)w$ & $\times(w+1)$ \\
\hline
\textbf{Graphicl State  }& $(nb+(nc+m)w)$ & $n_b+$ &$n_b+$\\
\textbf{Space model  }  &    $\times (n_b+n_c(w+1))$ & $(n_c+m)w$  &$n_c(w+1)$ \\
\hline
\end{tabular}
\label{tab1}
\end{center}
\end{table}
\par
Graphical state space model is illustrated with state space form in Fig. 7. It can be also represented in factor graph form as it is illustrated in Fig. 8. Obviously, it is a graphical model with cycles. This means that junction tree algorithm must be used to represent the model with a tree without cycles  for exact inference. The estimation will get convergent to a fixed point when
the probabilistic inference is being perfermed via the junction tree. It will be demonstrated in the next section that
graphical state space model can describe  the system more accurately than Extended Kalman filter in some cases.

\section{A simple example}
A simple radar tracking model was usually used to demonstrate how to use Kalman filter\cite{b15}. Here it is used to demonstrate the efficiency of the new framework.
However, it should be noted that this framework can not be used in all kinds of system!!! For this reason, it is strongly recommended that
you must perform simulation before real implementation.
\par
A simple two dimensional radar-tracking problem can be described as follows
\begin{gather}
X=\begin{bmatrix}
distance\\
velocity\\
altitude
\end{bmatrix}=
\begin{bmatrix}
x \\
\dot{x} \\
h
\end{bmatrix}\notag
\\
\dot{x}=0
\end{gather}
The measured range
\begin{gather}
\rho=\sqrt{x^2+h^2}
\end{gather}
The continous-time system of simple radar tracking is described by
\begin{gather}
\begin{bmatrix}
\dot{x}  \\
\ddot{x} \\
\dot{h}
\end{bmatrix}
=\begin{bmatrix}
0 & 1  & 0 \\
0 & 0 & 0 \\
0 & 0 &0
\end{bmatrix}
\begin{bmatrix}
x  \\
\dot{x}\\
h
\end{bmatrix}+q\\
\tilde{\rho}
=\begin{bmatrix}
\frac{x}{\sqrt{x^2+h^2}}& 0 & \frac{h}{\sqrt{x^2+h^2}}
\end{bmatrix}\begin{bmatrix}
x  \\
\dot{x}\\
h
\end{bmatrix}
+r
\end{gather}
When the framework of Kalman filter is used,  the discrete-time system goes as follows
\begin{gather}
\begin{bmatrix}
x_{k+1} \\
\dot{x}_{k+1}\\
h_{k+1}
\end{bmatrix}
=\begin{bmatrix}
1 & T  & 0 \\
0 & 1 & 0 \\
0 & 0 &1
\end{bmatrix}\begin{bmatrix}
x_k  \\
\dot{x}_k\\
h_k
\end{bmatrix}
+q_k\\
\tilde{\rho}_{k+1}
=\begin{bmatrix}
\frac{x_{k+1}}{\sqrt{x^2_{k+1}+h^2_{k+1}}}& 0 & \frac{h_{k+1}}{\sqrt{x^2_{k+1}+h^2_{k+1}}}
\end{bmatrix}
\begin{bmatrix}
x_{k+1}  \\
\dot{x}_{k+1}\\
h_{k+1}
\end{bmatrix}
+r_{k+1}
\end{gather}
where $T$ is the prediction interval.
\par If graphical state space model is used, the system state can be  modeled distributedly.
\begin{gather}
\alpha_{k+1}=\frac{x_{k+1}}{\sqrt{x^2_{k+1}+h^2}},\beta_{k+1}=\frac{h}{\sqrt{x^2_{k+1}+h^2}}
\end{gather}
The discrete-time system is discribed by
%\begin{figure*}[hb]
%\begin{align}
\begin{gather}
A_w=
\begin{bmatrix}
1  & 0 & 0 & 0 & 0  &\cdots & 0 & 0 \\
0 & 1 & 0 & 0 & 0  &\cdots & 0 & 0 \\
0 & 0 & 1 & 0 & 0 &\cdots & 0 & 0 \\
0 & -T & -1& 1& 0 &\cdots &0 & 0 \\
\beta_{k+1} & 0& 0& \alpha_{k+1}& 0 &\cdots & 0 & 0 \\
0   &-T & 0& -1 &1 &\cdots & 0 & 0\\
\beta_{k+2} & 0&  0 & 0 & \alpha_{k+2}&\cdots &  0 & 0\\
\vdots &\vdots &\vdots  &\vdots  &\ddots & \vdots & \vdots& \vdots \\
0& -T& 0& 0 &0 & \cdots &-1 & 1\\
\beta_{k+w} & 0 &0 &0&0&\cdots &0 &  \alpha_{k+w}
\end{bmatrix}
\end{gather}
%\end{align}
%\end{figure*}

\begin{gather}
  P_w=diag
\begin{bmatrix}
P_{h}\\
P_{\dot{x}}\\
P(k)\\
Q(k)\\
R_{k+1}\\
Q(k+1)\\
R_{k+2}\\
\vdots\\
Q(k+w-1)\\
R_{k+w}\\
\end{bmatrix},
X_w=
\begin{bmatrix}
  h  \\
  \dot{x}\\
  x_k\\
  x_{k+1}\\
  x_{k+2}\\
  \vdots \\
  \vdots \\
  x_{k+w-1}\\
  x_{k+w}
  \end{bmatrix},
\notag
\end{gather}
\begin{gather}
b_w=
\begin{bmatrix}
\hat{h}^+  \\
\hat{\dot{x}}^+ \\
\hat{x}^+(k)  \\
0\\
y_{k+1}\\
0\\
y_{k+2}\\
\vdots \\
0\\
y_{k+w}\\
\end{bmatrix},\notag
\\
A_wX_w=b_w
\end{gather}
For Kalman filter, simulation settings are set as follows
\begin{gather}
    q_k \sim N(0,Q_k),Q_k=
  \begin{bmatrix}
  0.005^2 & 0 & 0\\
  0  & 0.005^2 &0\\
  0& 0 &0
  \end{bmatrix}\notag
  \\
  r_k \sim N(0,R_k),R_k=9\notag
  \\
T=0.05,P_0=\begin{bmatrix}
49 & 0 & 0\\
0  & 49 & 0\\
0  & 0  & 49
\end{bmatrix},x_0=\begin{bmatrix}
-100\\
200\\
2000
\end{bmatrix}
\end{gather}
For graphical state space model, simulation settings are set as follows
\begin{gather}
  T=0.05,Q_k=0.005^2,R_k=9, \notag
  \\
P_h=49,P_{dot}=49,P(0)=49,\notag
\\
x^+_0=-100,\dot{x}^+_0=200,h^+=2000,w=10
\end{gather}
Obviously, Kalman filter and graphical state space model adopt the same simulation settings.
Simulation results\footnote{The open soure codes for this example can be found at github,  https://github.com/shaolinbit/GraphicalStateSpaceModel} are demonstrated in Fig. 9 and Fig. 10. For this example, graphical state space model can  outperform Kalman filter.
\begin{figure}[htbp]
\centerline{\includegraphics[scale=0.3]{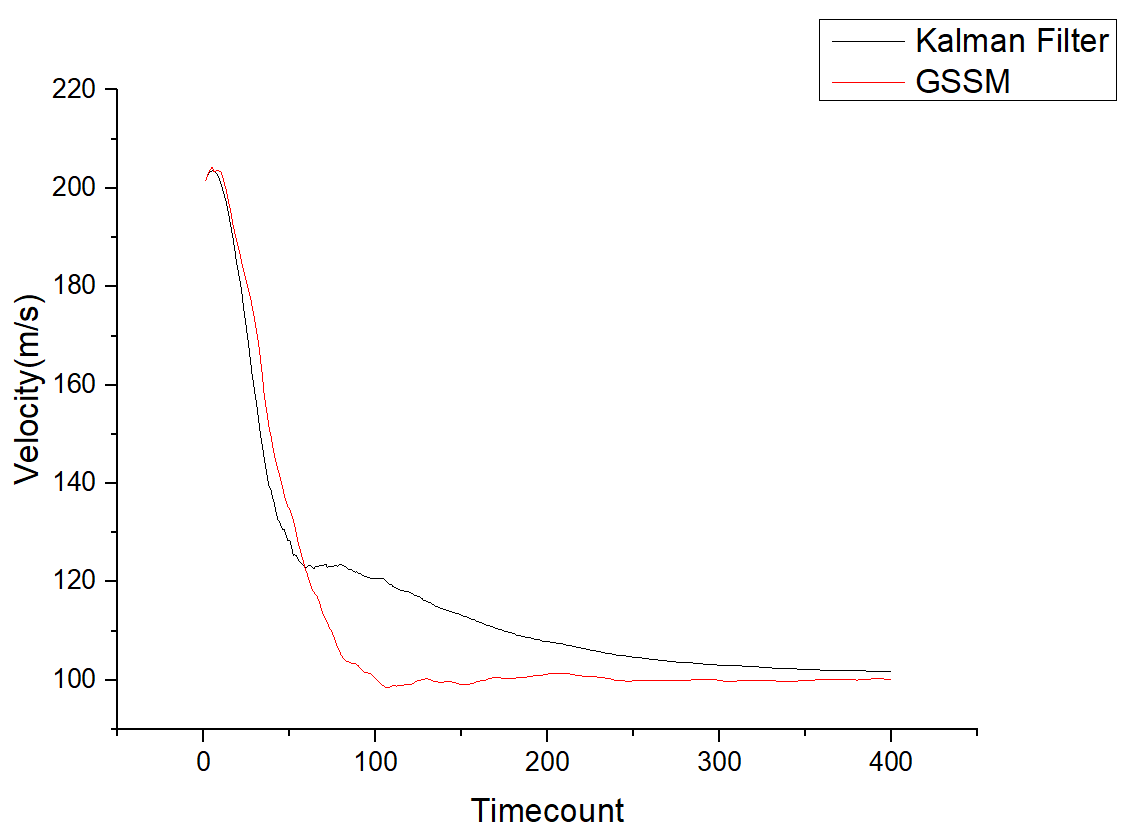}}
\caption{Velocity Estimation.}
\label{fig}
\end{figure}

\begin{figure}[htbp]
\centerline{\includegraphics[scale=0.28]{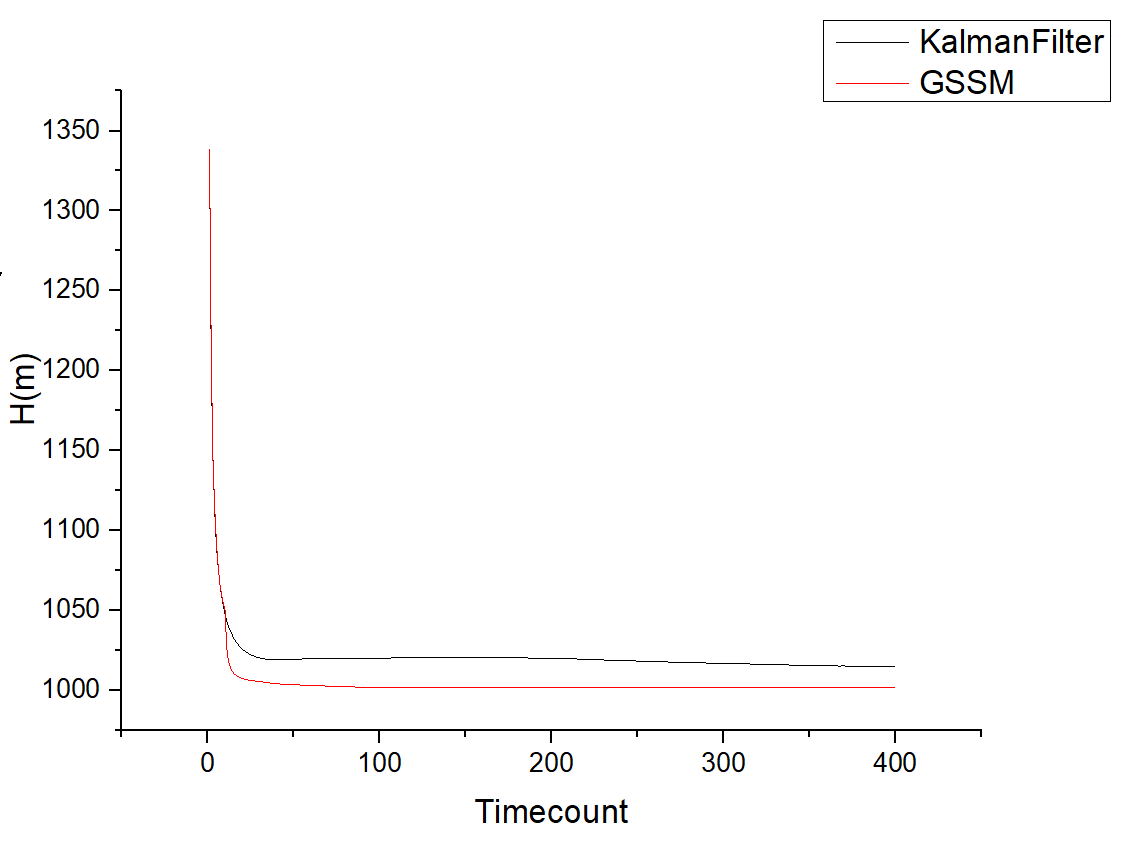}}
\caption{Estimation of H.}
\label{fig}
\end{figure}

\section{Conclution and Future works}
In this paper, a new framework, named as graphical state space model, was proposed to estimate the state of a class  of nonlinear system.
A simple example was used to demonstrate the efficiency of this framework.
In some cases, graphical state space model can extract more information than the standard Kalman disrete model.
Detailed error analysis should be done in the future.
 What should be explorered next is that which kind of system can
  be estimated by this framework.

\section*{Acknowledgment}
Graphical state space model\cite{b16}  is  patent pending.
Thanks for Professor Yulong Huang from Harbin University of Engineering for his kindly invitation.

\section*{}

\vspace{12pt}
\color{red}
%IEEE conference templates contain guidance text for composing and formatting conference papers. Please ensure that all template text is removed from your conference paper prior to submission to the conference. Failure to remove the template text from your paper may result in your paper not being published.

\end{document}